\documentclass[12pt]{article}
\usepackage{amsmath}
\usepackage{amssymb}
\usepackage{amsthm}
\usepackage{bm}
\usepackage{bbm}
\usepackage{upgreek}
\usepackage{graphicx}
\usepackage{enumerate}
\usepackage{enumitem}
\usepackage{natbib}
\usepackage{url}
\usepackage{hyperref}
\usepackage[english]{babel}
\usepackage{epstopdf}
\usepackage{tabularx}
\usepackage{float}
\usepackage{multirow}
\usepackage{algorithm,algpseudocode}

\newtheorem{theorem}{Theorem}
\newtheorem{assumption}{Assumption}

\newtheorem{proposition}{Proposition}

\newtheorem{definition}{Definition}

\newcommand{\blind}{0}

\newcommand{\indep}{\perp \!\!\! \perp}
\newcommand{\X}{{\bf X}}
\newcommand{\x}{{\bm{x}}}
\newcommand{\z}{{\bm{z}}}

\newcommand{\Z}{{\bf Z}}
\newcommand{\Y}{{\bf Y}}

\newcommand{\B}{{\bf B}}
\newcommand{\D}{{\bf D}}

\newcommand{\ba}{{\bf a}}

\newcommand{\spc}{{\mathcal S}_{Y|\X}}

\newcommand{\cms}{{\mathcal S}_{E(Y|\X)}}

\newcommand{\R}{\mathbb{R}}
\newcommand{\mS}{\mathcal{S}}

\newcommand{\A}{{\bf A}}

\newcommand{\M}{{\bf M}}

\newcommand{\g}{{\bf g}}

\newcommand{\bP}{{\bf P}}

\newcommand{\bbeta}{\bm{\beta}}

\newcommand{\var}{\text{Var}}
\newcommand{\Sig}{\bm{\Sigma}}
\newcommand{\bmu}{\bm{\mu}}
\newcommand{\cov}{\text{Cov}}
\newcommand{\bOmega}{\bm{\Omega}}
\newcommand{\bGamma}{\bm{\Gamma}}
\newcommand{\balpha}{\bm{\alpha}}

\newcommand{\I}{{\bf I}}

\DeclareMathOperator*{\argmin}{arg\,min}
\DeclareMathOperator{\E}{E}

\begin{document}

\def\spacingset#1{\renewcommand{\baselinestretch}%
{#1}\small\normalsize} \spacingset{1}

%%%%%%%%%%%%%%%%%%%%%%%%%%%%%%%%%%%%%%%%%%%%%%%%%%%%%%%%%%%%%%%%%%%%%%%%%%%%%%

\if0\blind
{
  \title{\bf Sufficient dimension reduction for regression with metric space-valued responses}
  \author{Abdul-Nasah Soale\thanks{
    Corresponding author: \textit{abdul-nasah.soale@case.edu}}\hspace{.2cm}\\
    Department of Mathematics, Applied Mathematics, and Statistics\\ Case Western Reserve University, Cleveland, OH, USA\\
    and \\
    Yuexiao Dong \\
   Department of Statistics, Operations, and Data Science \\ Temple University, Philadelphia, PA, USA}
  \maketitle
} \fi

\if1\blind
{
  \bigskip
  \bigskip
  \bigskip
  \begin{center}
    {\LARGE\bf Title}
\end{center}
  \medskip
} \fi

\bigskip
\begin{abstract}
\noindent Data visualization and dimension reduction for regression between a general metric space-valued response and Euclidean predictors is proposed. Current Fr\'ech\'et dimension reduction methods require that the response metric space be continuously embeddable into a Hilbert space, which imposes restriction on the type of metric and kernel choice. We relax this assumption by proposing a Euclidean embedding technique which avoids the use of kernels. Under this framework, classical dimension reduction methods such as ordinary least squares and sliced inverse regression are extended. An extensive simulation experiment demonstrates the superior performance of the proposed method on synthetic data compared to existing methods where applicable. The real data analysis of factors influencing the distribution of COVID-19 transmission in the U.S. and the association between BMI and structural brain connectivity of healthy individuals are also investigated. %The theoretical justifications are included as well.
\end{abstract}

\noindent%
\vfill

\newpage
\spacingset{1.75} % DON'T change the spacing!
\section{Introduction}
\label{sec:intro}
\noindent 
Regression analysis with Euclidean data $(\Y,\X)$, where the response $\Y\in\R^q$ and predictor $\X\in\R^p$, for some positive integers $p,q\geq 1$, is well-established. Lately, however, there has been a growing demand for regression with general metric space-valued objects that may belong to non-Euclidean spaces such as the space of positive definite matrices, probability distributions, networks, and spheres, to mention a few. Examples of regression with such random objects can be found in \cite{faraway2014regression, petersen2019frechet}, and \cite{petersen2019frechettime}. In some of these non-Euclidean metric spaces, standard properties of vector spaces such as inner products may not be applicable, which makes it difficult to implement classical regression methods directly. To address this problem, Fr\'ech\'et regression was recently proposed by \cite{petersen2019frechet}.

Fr\'ech\'et regression basically extends the classical regression in the Euclidean space to responses in general metric spaces. 
Let $(\bOmega_Y, \rho)$ denote a metric space equipped with a metric $\rho$. For a random object $(\X,Y) \in (\R^p\times \bOmega_Y)$, assume the conditional distributions $F_{X|Y}$ and $F_{Y|X}$ are well-defined. Then the conditional Fr\'ech\'et mean of $Y|\X=\x$ is given by
\begin{align}
    m_{\oplus}(\x) = \argmin_{\omega\in\bOmega_Y} \E[\rho^2(Y,\omega)|\X=\x].
    \label{frechet_reg}
\end{align}
From model (\ref{frechet_reg}), it is not hard to see that beyond the complex nature of $Y\in\bOmega_Y$, the problems associated with $\X$ in the classical mean regression such as large $p$ and outliers, easily carryover to Fr\'ech\'et regression. Thus, sufficient dimension reduction still plays a crucial role in visualization and extracting relevant information in Fr\'ech\'et regression with high dimensional $\X$. 

For the regression between $Y\in\bOmega_Y$ and $\X\in\R^p$, a subspace $\B\in\R^{p\times d}$ is a Fr\'echet dimension reduction subspace if
\begin{align}
	Y \indep \X |\B^\top\X,
	\label{sdr}
\end{align}
where $``\indep"$ means statistical independence and $\B^\top\B = \I_d$, with $d < p$. While $\B$ is not identifiable, it can be shown that the column space of $\B$, denoted as $\mathrm{span}(\B)$, is unique under some mild conditions (\cite{cook1996graphics} and \cite{yin2008successive}). If it exists, the intersection of all $\B$'s that satisfies (\ref{sdr}) will have the smallest column space and is called the Fr\'echet central space for the regression of $Y$ on $\X$. We denote the central space as $\spc=\mathrm{span}(\B)$ and $dim(\spc) = d$ as the structural dimension.

Sometimes we may only be interested in some part of the central space such as the mean space. In such a case, we restrict our attention to finding $\bGamma \in\R^{p \times d}$ such that 
\begin{align}
    Y \indep \E(Y|\X) | \bGamma^\top\X,
    \label{cms}
\end{align}
where $\bGamma^\top\bGamma = \I_d$. The smallest column space of $\bGamma$ for which (\ref{cms}) holds is called the central mean space denoted as $\cms$. The central mean space is a subspace of the central space, i.e., $\cms \subseteq \spc$. Hence, $\mathrm{span}(\bGamma) \subseteq \mathrm{span}(\B)$.

Recently proposed Fr\'echet dimension reduction methods can be found in the works of \cite{ying2022frechet, dong2022frechet}, and \cite{zhang2023dimension}. The central theme in these methods involves finding an equivalent representation of $Y$ in the Hilbert space that preserves the distance between the original objects in $\bOmega_Y$. With the embedding, existing dimension reduction techniques such as ordinary least squares (OLS; \cite{li1989regression}), sliced inverse regression (SIR; \cite{li1991sliced}), minimum average variance estimation (MAVE) and its variants (eg. \cite{xia2002mave, xia2009adaptive}) are then applied. It is assumed that $(\bOmega_Y,\rho)$ is continuously embeddable in a Hilbert space to allow for the construction of a {\it universal kernel} on $\bOmega_Y$. 

However, this comes with some restrictions on the metric space. For instance, $(\bOmega_Y,\rho)$ needs to be of the negative type to allow for embedding in a complete and separable Hilbert space as stated in Proposition 2 of \cite{zhang2023dimension}. While this assumption may be satisfied in many applications, the requirement can be quite restrictive. Moreover, whether a given kernel is universal or not depends on the metric space under consideration, as not all popular kernels such as the Gaussian and Laplacian kernels are universal beyond the Euclidean space, see \cite{micchelli2006universal}. Therefore, choosing the appropriate kernel is crucial even if the metric space is continuously embeddable in a Hilbert space. Besides, constructing kernels come with the additional cost of choosing the optimal tuning parameters, which is not trivial in some cases. Lastly, using the kernel approach can be computationally intensive. 

In this paper, we propose a Euclidean embedding approach which allows for slight distortions. By merely requiring $(\bOmega_Y,\rho)$ to be embeddable in the Euclidean space, we avoid the continuous embedding condition. Note that the metric $\rho$ already maps any two objects in the response space to $\R$. Thus, we only need to find a low dimensional representation of the pairwise distance matrix, which preserves the distance between the objects in their original space. There is an extensive body of work on how to do this, see \cite{huber1985projection, bickel2018projection, matousek2013}, and \cite{li2008projective}. We adopt the random projections technique, which is easier to implement and computationally faster. In addition, complications associated with the kernel approach including choosing the appropriate kernel for specific metrics and finding optimal tuning parameters are avoided.

The rest of the paper is organized as follows. In Section 2, we review some concepts on embedding between metric spaces. The ensemble estimates of the central space at the population level are discussed in Section 3. In the Section 4, we present a few metric choices for specific metric spaces and the estimation algorithm for the central space. The performance of our proposed surrogate-assisted techniques compared to existing methods are provided for synthetic and real data in Sections 5 and 6, respectively. The paper concludes with a discussion in Section 7. All proofs are relegated to the appendix.

\section{Embedding metric spaces into Euclidean space}
This section reviews some properties of metric spaces.

\begin{definition}[Metric]
For a non-empty set $\bOmega_Y$, let the function $\rho:\bOmega_Y \times \bOmega_Y \to [0,\infty)$. Then $(\bOmega_Y,\rho)$ is a metric space equipped with a metric $\rho$ if for all $y,y',y^* \in \bOmega_Y$, the following properties are satisfied: (1) positivity: $\rho(y,y') \geq 0$ with equality if and only if $y=y'$; (2) symmetry: $\rho(y,y') = \rho(y',y)$; and (3) triangle inequality: $\rho(y,y') \leq \rho(y,y^*) + \rho(y^*,y')$.

\end{definition}

\noindent Therefore, for a finite subset of points $\{Y_1,\ldots,Y_n \}\in \bOmega_Y, n \geq 2$, we can compute the pairwise distance matrix $\D_n \in\R^{n\times n}$ with elements $(\D_n)_{ij}=\rho(Y_i,Y_j)$, for $i,j=1,\ldots,n$. 

\begin{definition}[Isometric embedding]
Let $(\bOmega_Y,\rho_Y)$ and $(\bOmega_z,\rho_Z)$ be metric spaces. A mapping $f:(\bOmega_Y,\rho_Y) \to (\bOmega_Z,\rho_Z)$ is an isometric embedding if $\rho_Y(y,y') = \rho_Z\big(f(y),f(y')\big)$ for all $y,y'\in\bOmega_Y$.
\end{definition}
In other words, an isometric embedding exists if $f(.)$ is a bijection. For example, let $\{Y_1,\ldots,Y_n\}\in \bOmega_Y$ and  $\{\ba_1,\ldots,\ba_n\} \in \R^k, 1\leq k \leq n$. Then $\{\ba_1,\ldots,\ba_n\} $ is an isometric embedding of $\{Y_1,\ldots,Y_n\}$ if $\rho (Y_i,Y_j)=\lVert a_i - a_j \rVert_2 $,  $\forall i,j=1,\ldots,n$, where $\lVert.\rVert_2$ denotes the Euclidean norm. By theorem 2.4 of \cite{wells1975embeddings}, $(\bOmega_Y,\rho)$ is isometrically embeddable in a Hilbert space if $(\bOmega_Y,\rho)$ is of a negative type. This means, for every finite subset $\{Y_1,\ldots,Y_n\}\in\bOmega_Y, n\geq 2$, there exists $\alpha_1,\ldots,\alpha_n \in \R$ with $\sum_{i=1}^n\alpha_i =0$ such that the quadratic form
\begin{align}
    \sum_{i=1}^n\sum_{j=1}^n \alpha_i\alpha_j\rho(Y_i,Y_j) \leq 0.
\end{align}

While isometric embeddings may exist theoretically, it is often impossible to represent one metric space exactly into another. Moreover, for the purpose of visualization and recovering unbiased estimates of the central space or central mean space, an approximate embedding with slight {\it distortion} may suffice. 

\begin{definition}[$\epsilon$-almost isometry]
An $\epsilon$-almost isometry between the metric spaces $(\bOmega_Y,\rho_Y)$ and $(\bOmega_z,\rho_Z)$ exists if for some $\epsilon \in (0,1)$, a mapping $f:(\bOmega_Y,\rho_Y) \to (\bOmega_z,\rho_Z)$ is such that 
\begin{align}
    (1-\epsilon)\rho_Y(y,y') \leq \rho_Z\big(f(y),f(y')\big) \leq (1+\epsilon)\rho_Y(y,y'), \ \text{for all} \ y,y'\in\bOmega_Y.
\end{align}
\end{definition}

\noindent In the Euclidean space, such approximate embeddings can be obtained through linear maps. 

\begin{theorem}[Johnson–Lindenstrauss Lemma]\label{JL}
Let $\epsilon \in (0,1)$ and for any integer $n$, let $k$ be a positive integer such that $k \geq C\epsilon^{-2}\log n$. Then for any set of $n$ points
$\{\ba_1,\ldots,\ba_n\} \in \R^p$, there is a mapping 
$f:\R^p \to \R^k$ such that 
\begin{align*}
    (1-\epsilon)\lVert \ba_i-\ba_j \rVert_2 \leq \lVert f(\ba_i)-f(\ba_j) \rVert_2 \leq (1+\epsilon)\lVert \ba_i-\ba_j \rVert_2, \forall i,j=1,2,\ldots,n. 
\end{align*}
\end{theorem}

Theorem \ref{JL} is popular in the dimension reduction literature. It implies that every high dimensional Euclidean space can be embedded in a low dimensional Euclidean space with dimension in $\log(n)$. In fact, $f$ is a random linear map by Lemma 2.5.2 of \cite{matousek2013}. It has also been shown that projecting the high dimensional Euclidean space onto a large finite set of unit vectors on the hypershpere $\mathbb{S}^{p-1}$ achieves an $\epsilon$-almost isometry. See \cite{matousek2013}.

\section{Ensemble estimates of the central space}
\subsection{General background to the ensemble technique}
By the definition of sufficient dimension reduction in (\ref{sdr}), if we let 
$\widetilde{Y}$ be a random copy of $Y$, then $Y \indep \X |  \B^\top\X  \implies  (Y, \widetilde{Y}) \indep \X | \B^\top\X$. Also, by the transformational properties of the central space (Theorem 2.3 of \cite{li2018sufficient}), $(Y,\widetilde{Y}) \indep \X |  \B^\top\X  \implies  \psi(Y, \widetilde{Y}) \indep \X | \B^\top\X$, for any measurable function $\psi$. Therefore, we can define a surrogate response $Y^s$ as a mapping $f:\rho(y,y') \to \R$, for all $y,y'\in \bOmega_Y$, where $f(.)$ is some square integrable linear function. 

\begin{proposition}\label{surr_sdr}
$\mathcal{S}_{Y^s|\X} \subseteq \spc$ and $\mathcal{S}_{\E(Y^s|\X)} \subseteq \cms$. 
\end{proposition}

\noindent Proposition \ref{surr_sdr} implies that both $\spc$ and $\cms$ can be characterized using a collection of subspaces for the regression of $Y^s$ versus $\X$. As proposed by \cite{yin2011sufficient}, if we choose a rich collection of functions $Y^s$, we can estimate the central space by aggregating the mean subspaces for each function. Here, we extend this idea to include aggregating the central subspaces as well.

\begin{definition}\label{def_ensemble}
    Let $\xi$ be a family of functions $\psi: \Omega_Y \to \R$. Then $\xi$ characterizes
    \begin{align}
    \text{ (a) } & \cms \text{ if } \mathrm{span} \big\{\mS_{\E(Y^s|\X)}: \psi \in \xi \big\} = \cms. \\
    \text{ (b) } & \spc \text{ if }\mathrm{span} \big\{\mS_{Y^s|\X}: \psi \in \xi \big\} = \spc
    \end{align}
\end{definition}

\noindent Part (a) of Definition \ref{def_ensemble} is called the central mean space (CMS) ensemble while (b) is referred to as the central space (CS) ensemble. We will denote either the CMS or CS ensemble for $\psi \in \xi$ as $\mS(\xi)$. 

\begin{proposition}\label{ensemble_sdr}
Let $L_2(F_Y)$ be the class of square-integrable functions of $Y^s$ and $\mathcal{I}$ be a class of measurable indicator functions of $Y^s$. Also assume $L_2(F_Y)$ is dense in $\mathcal{I}$. Then $\xi \subseteq L_2(F_Y)$ implies that $\mS(\xi ) \subseteq \cms \subseteq \spc$, for all $\psi \in \xi$.
\end{proposition}

Following Proposition \ref{ensemble_sdr}, our ensemble estimate of $\spc$ can be obtained from the spectral decomposition of the candidate matrix $\M: P_{XY} \times \xi \to \R^{p \times p}$ such that $\mathrm{Span}(\M) \subseteq \spc$, where $P_{XY} = \{F_{XY} \}$ is a collection of all distributions of $(\X,Y)$. Note that some conditions required for the original estimator also carryover to the ensemble case. For instance, the linear conditional mean (LCM) assumption is still required for the Fr\'echet central space estimates based on OLS and SIR ensembles as LCM is imposed on only the predictor.

\subsection{Surrogate-assisted OLS and SIR estimators of $\spc$}
The original OLS and SIR both require the linear conditional mean (LCM) assumption and applies here as well.
\begin{assumption}[LCM]
 We assume $\E(\X|\B^\top\X)$ is a linear function of $\B^\top\X$ if $\mathrm{span}(\B) \subseteq \spc$. 
\end{assumption}
We start with the surrogate-assisted OLS (sa-OLS) estimator. First, we assume $\X$ satisfies the LCM assumption and that $\Sig = \var(\X)$ is nonsingular. Next, for all $\psi\in\xi$, let
\begin{align}
   \beta( \psi) &= \Sig^{-1}\Sig_{XY^s} \in \spc, \\
   \M_{ols} &= \E\big(\beta( \psi)\beta^\top(\psi)\big),
   \label{ols_ensem}
\end{align}
where $\Sig_{XY^s}$ denotes the covariance between $\X$ and $Y^s$ and the expectation in (\ref{ols_ensem}) is taken over all $\psi\in\xi$. Thus, $\mathrm{span}(\M_{ols}) \subseteq \spc$. Notice that unlike the classical OLS, the sa-OLS estimator can recover more than one direction in the central space. Moreover, for Euclidean responses, sa-OLS can handle heteroscedastic error models with monotone links even when $\E(Y|\X) = \{0\}$. The latter will be demonstrated in the simulation study. 

We now turn  our attention to the surrogate-assisted sliced inverse regression (sa-SIR) estimator. As in the sa-OLS case, we assume $\X$ satisfies the LCM assumption and $\Sig = \var(\X)$ is nonsingular. Then the candidate matrix for the sa-SIR estimator is given by
\begin{align}
    \M_{sir}(., \psi) = \E\left\{\var\big[\E\big(\X|Y^s\big)\big]\right\},
\end{align}
where the outer expectation $\E\{.\}$ is taken over all $\psi\in\xi$.

\section{Metric choice and estimation algorithm}
Let $(\x_1, y_1),\ldots,(\x_n,y_n)$ be a random sample of $(\X,Y)$. Following Theorem \ref{JL}, we define $y^s = u^\top\D_n$, where $\D_n$ is a matrix of pairwise distances and $u\in \mathbb{S}^{n-1}:\lVert u \rVert_2=1$. It is worth noting that for any given metric space, there may be several choices for $\rho$. However, we will only focus on popular metrics used in the existing literature for the response metric spaces considered here.

\subsection{Metric choice for different response spaces}
\subsubsection{Euclidean responses}
While the plethora of SDR methods are design for the Euclidean response, it is still worth exploring how the proposed method fares in this space, at least for a start, as this space is more tractable. 
\noindent Let $Y_n=\{y_1, \ldots, y_n\} \in \bOmega_Y \subseteq  \R^q$, where $q \geq 1$. For $q=1$, we estimate the pairwise distances as $(\D_n)_{ij} = |y_i-y_j|, \forall i,j=1,\ldots,n$. When $q > 1$, we propose to use the $\ell_2$-norm: $(\D_n)_{ij} = \big( \sum_{k=1}^q (y_{ik}-y_{jk})^2 \big)^{1/2}$. The Mahalanobis distance is arguably more appropriate for high dimensional responses if it is suspected that the correlation between the responses is associated with the predictor.

\subsubsection{Distribution as response}
If $\bOmega_Y$ is a space of probability distributions, we propose to use the Wasserstein metric as used in \cite{petersen2019frechet, dong2022frechet}, and \cite{zhang2023dimension}). The Wasserstein metric for probability measures $\zeta$ and $\nu$ on $\R^q$ with finite $k$th moment is given by 
\begin{align}
	W_k(\zeta,\nu) = \inf_{\substack{y \sim \zeta \\ y'\sim \nu}} \big (\E\lVert y - y'\rVert_k \big )^{1/k}, \quad \quad k \geq 1,
	\label{wasserstein}
\end{align}
where $\lVert . \rVert_k$ is the $\ell_k$-norm, $y$ and $y'$ marginally distributed as $\zeta$ and $\nu$, respectively, and the infimum is taken over all pairs of $(y, y') \in \R^q$ (\cite{panaretos2019statistical})). 

For univariate distributions, $W_k(.)$ is explicitly defined as 
\begin{align}
 W_k(y,y') = \lVert F_{y}^{-1} - F_{y'}^{-1}\rVert_k = \left(\int_0^1 \big\lvert F_{y}^{-1}(s)  - F_{y'}^{-1}(s)\big\rvert^k ds \right )^{1/k},
\label{quad_wasserstein}
\end{align}
where $F_{y}^{-1}$ and $F_{y'}^{-1}$ are the respective quantile functions of $y$ and $y'$. We prefer $W_1(.)$ over the quadratic Wasserstein ($W_2$) used in previous studies because the former is more robust to outliers.

\subsubsection{Network as response}
Consider a network or graph $G=(V,E)$, where $V$ is the set of {\it vertices (nodes)} and $E$ is the set of {\it edges (links)}. The {\it order} of $G$ is   $N_V=|V|$ and its {\it size} is $N_E = |E|$. Two distinct nodes $v_1, v_2 \in V$ are {\it connected } if there is an edge between them, denoted as the ordered pair $\{v_1, v_2 \}$. The {\it neighborhood} of node $v_1$ is given by $\Gamma(v_1) = \big\{v_j \in V: \{v_1,v_j \}\in \E, v_1 \neq v_j \big\}$. The {\it degree} of $v_1$ is $|\Gamma(v_1)|$. 

There are several ways to characterize the topological structure of a network. A common approach is to use the {\it centrality} measure. Centrality $C$ is a function defined over all nodes that returns a positive value, i.e., $C: (G, v) \to [0, \infty)$, for all $v \in V$. A centrality measure can be specific to the degree, closeness, betweenness, or eigenvectors of the adjacency matrix. Let $G_1=(V,E_1)$ and $G_2=(V,E_2)$ be two random networks of the same order. The centrality distance (CD) proposed by \cite{roy2014modeling} is given by 
    \begin{align}
        d_C(G_1,G_2) = \displaystyle\sum_{v\in V} |C(G_1,v) - C(G_2,v)|.
        \label{cent_dist}
    \end{align}
Our centrality measure of choice is the degree centrality (node degree), which in effect represents some notion of ``importance" or popularity of the node. Degree centrality distance does not account for the weight of the link between the nodes, which is very important in weighted networks. 

To account for edge weights, we may use the {\it graph diffusion distance (DD)} proposed by \cite{hammond2013graph}. This distance measure involves two steps. First, we find the graph Laplacian exponential kernel matrix of each graph and then take the Frobenius norm of the difference of kernels. Let $\A_{G_1}$ and $\A_{G_2}$ be the adjacency matrices of graphs $G_1$ and $G_2$, respectively. The corresponding degree matrix of graph $G_i$ is given by $\D_{G_i}$, which is an $N_V \times N_V$ diagonal matrix whose diagonal entries are $(\D_{G_i})_{jj} = \displaystyle\sum_{k=1}^{N_V} (\A_{G_i})_{jk}$. The corresponding Laplacians are given by $\L_{G_i} = \D_{G_i} - \A_{G_i}$. Then, at any given time $t$, the diffusion distance is given by
\begin{align}
    d_t(G_1,G_2) = \max_t \big(\lVert e^{-t\L_1} - e^{-t\L_2}\rVert_F^2\big)^{1/2},
\label{diff_dist}
\end{align}
where $\lVert . \rVert_F$ is the matrix Frobenius norm. For static networks, we fixed $t=1$.

\subsubsection{Functional responses}
\noindent Here, we consider responses measured over time but not necessarily at fixed time intervals. A good metric for this space  is one that takes into account both the phase and amplitude variations. Based on this idea, \cite{faraway2014regression} proposed to use the Fr\'echet distance. We choose to use the discrete Fourier distance of \cite{agrawal1993efficient}. For the discrete Fourier distance, we use the discrete Fourier transform (DFT) to map the response sequence to a low dimensional space to reveal the periodicities and their relative strengths at the periodic components. Then we find  the Euclidean distance between the modulus of the Fourier coefficients.

Let $y(t)$ denote the functional response measured at time (sample)  $t=1,\ldots, T$. The DFT of $y(t)$ is a sequence of complex numbers of the same length $T$ given by
\begin{equation}
    Y(k) = \displaystyle\sum_{t=0}^{T-1} y(t)e^{-i 2\pi kt/T}, k=0,1,2,\ldots,T-1,
\end{equation}
where $Y(k)$ denotes the $k$th spectral sample and $i=\sqrt{-1}$ is the imaginary unit. By Parseval's theorem, the Fourier transform preserves the Euclidean distance in both the time and frequency domains. Moreover, the Fourier transform maps the sequence to a low-dimensional space using only the first few Fourier coefficients, which means we do not need all the coefficients to reconstruct the original sequence.  Thus, we can find the distance between any two functional responses by calculating the Euclidean distance between the modulus of their  $\lfloor{T/2}\rfloor +1$ coefficients as used in the R package {\it TSdist} (\cite{tsdist}).

%The advantage of the DFT is that we can reconstruct the original sequence as follows:
% \begin{equation}
%     y(t) = \cfrac{1}{T}\displaystyle\sum_{k=0}^{T-1} Y(k)e^{i 2\pi kt/T}, k=0,1,2,\ldots,T-1.
% \end{equation}

\subsection{Estimation algorithms}
For ease of computation, it is a common practice to estimate the candidate matrix based on a standardized predictor $\Z = \Sig^{-1/2}(\X-\bmu)$, where $\bmu$ and $\Sig$ denote the mean and covariance of $\X$, respectively. By the invariance property of the central space, if $\M$ is the candidate matrix for $Y$ versus $\Z$, then $\Sig^{-1/2}\mathrm{Span}(\M) \in \spc$. The sa-OLS and sa-SIR estimation algorithms are provided in 1 and 2.

The procedure in Algorithm 2 can be followed to estimate the candidate matrix for other inverse regression methods not covered here such as sliced average variance estimation (SAVE; \cite{cook1991sliced}) and directional regression (DR; \cite{li2007directional}). While contour regression (\cite{li2005contour}) is not explored in this paper, the interested reader could use the computed distances directly to select the contour empirical directions or follow Algorithm 2.

\begin{algorithm}[htb]
\caption{sa-OLS}
\begin{algorithmic}[1]
   \State  Input: Predictor $\X$ as $(n \times p)$ matrix, response $Y_n = \{y_1,\ldots,y_n\}$ as list
   \State Define: $n\times n$ matrix $\D_n$, $(\D_n)_{ij} \gets d(y_i, y_j), \ \forall i,j = 1\ldots, n$
   \State Compute: $\overline{\x} = n^{-1}\displaystyle\sum_{i=1}^n \x_i$, $\hat\Sig =  n^{-1}(\x_i-\overline{\x})(\x_i-\overline{\x})^{\top}$, and $\hat\z \gets \hat\Sig^{-1/2} (\x-\overline{\x})$
   \State Set: $N \gets $ large integer ($\geq 1000$)
   \For{each $k$ in $1,\ldots,N$}
        \State $u_k \gets U/\lVert U \rVert$, $U \sim N(0,1)$
        \State $y^s_k \gets u_k^\top\D_n $
         \State $\hat\bbeta(u_k) \gets \hat\Sig^{-1}\hat\Sig_{zy^s_k}$, $\hat\Sig_{zy^s_k} = \cov(\hat\z, y^s_k)$
        \State $\M(u_k) \gets \hat\bbeta(u_k)\hat\bbeta^\top(u_k)$
    \EndFor
\State Compute: $\M_{ols}(U_N) \gets N^{-1}\displaystyle\sum_{1=1}^N \M(u_k)$, where $U_N=(u_1,\ldots,u_N)$. 
\State Return: $\hat{\B} = \hat\Sig^{-1/2}(\hat{\bm v}_1, \ldots, \hat{\bm v}_d)$, where $(\hat{\bm v}_1, \ldots, \hat{\bm v}_d)$ are the eigenvectors corresponding to the $d$ leading eigenvalues of $\hat \M_{ols}(U_N)$.
\end{algorithmic}
\end{algorithm}

\newpage

\begin{algorithm}[htb]
\caption{sa-SIR}
\begin{algorithmic}[1]
   \State  Repeat: 1-4 of Algorithm 1
   \For{each $k$ in $1,\ldots,N$}
        \State $u_k \gets U/\lVert U \rVert$, $U \sim N(0,1)$
        \State $y^s_k \gets u_k^\top\D_n $ 
        \State partition $supp(y^s_k)$ into intervals $I_1(u_k),\ldots, I_H(u_k)$. 
        \State Set: $I_{hi}(u_k) \gets \I (y^s_k \in I_h(u_k))$
        \State $\hat p_h(u_k) \gets n^{-1}\sum_{i=1}^n I_{hi}(u_k)$
	\State	$\hat \bmu_h(u_k) \gets \{n \hat p_h(u_k) \}^{-1}\sum_{i=1}^n \hat\z_iI_{hi}(u_k)$
        \State $\M(u_k) \gets \sum_{h=1}^H \hat p_h(u_k) \hat\bmu_h(u_k)\hat\bmu_h^\top(u_k)$
    \EndFor
  \State Compute: $\M_{sir}(U_N) \gets N^{-1}\sum_{1=1}^N \M(u_k)$  
 \State Return: $\hat{\B} = \hat\Sig^{-1/2}(\hat{\bm v}_1, \ldots, \hat{\bm v}_d)$, where $(\hat{\bm v}_1, \ldots, \hat{\bm v}_d)$ are the eigenvectors corresponding to the $d$ leading eigenvalues of $\hat\M_{sir}(U_N)$.
\end{algorithmic}
\end{algorithm}

\subsection{Asymptotic properties of the surrogate-assisted estimator}
The asymptotic behavior of the ensemble estimators depend on the terms of the number of projections $N$ and the sample size $n$. First, for any given $u$, $u^\top\D_n$ is can be viewed as a projection of $\D_n$ in the direction of $u$, which represents the characteristic function of $\D_n$ at $u$. When $n$ is large and $u$ is sparse, \cite{bickel2018projection} showed that $u^\top\D_n$ will be approximately normal. Thus, asymptotically, we expect the sa-OLS to be close to the best linear unbiased estimator if the regression link is monotone and $\D$ is sparse.

However, regardless of the asymptotic distribution of $u^\top\D_n$, the weak law of large numbers guarantees that  $\M(U_N) \to \E[\M(U)]$ as $N\to \infty$, where $U$ is a random vector uniformly distributed on $\mathbb{S}^{n-1}$. Also, since $\M(u_k)$ is positive semidefinite, $\M(U_N)$ takes each $\M(u_k)$ into account. Another thing to consider is that $\M_n(u)$ is a function of the moments based on the sample $(\x, u^\top\D_n)$, which can be expressed as a function of the empirical distribution $F_n(u)$. Hence, $\M_n(u)$ can be expanded around $F_0(u)$, the true distribution function of $(\X,Y)$ under some regularity conditions detailed in \cite{fernholz1983mises}.

Suppose Hadarmard differentiability holds and that $\M_n(u)$ can be expanded around $F_0(u)$ for any $u\in\mathbb{S}^{n-1}$. We further assume the following expression holds   
\begin{align}
	\M_n(u) = \M(u) + \E_n[\bm \Psi(\X,\D_n,u)] + \bm R_n(u),
	\label{taylor}
\end{align}
where $\M(u)$ is the true value whose estimate is $\M_n(u)$, $\bm \Psi(\X,\D_n,u)$ is a square integrable function statisfying the following:
\begin{align}
    \E[\bm \Psi(\X,\D_n,u)] &= \bm 0, \label{mean_cond}\\
    \sup_{u\in \mathbb{S}^{n-1}} \lVert \bm R_n(u) \rVert_F &= o_p(n^{-1/2}),
    \label{remainder_cond}
\end{align}
where $\E_n[\bm \Psi(\X,\D_n,u)]$ denotes the sample estimate of $\E[\bm \Psi(\X,\D_n,u)]$ and $\bm R_n(u)$ is a remainder term. When these conditions in (\ref{mean_cond}) and (\ref{remainder_cond}) are satisfied as well as the regularity conditions, $\M_n(U_N)$ is guaranteed to be $\sqrt{n}$-consistent for $\E[\M(U)]$ and asymptotically normal. See \cite{li2008projective} for more details.

\begin{theorem}\label{consistency1}\quad \\
Let $U$ be a random vector uniformly distributed on $\mathbb{S}^{n-1}$. Given a sample of size $n$ and $\bm U_N = (u_1,\ldots, u_N)$ for some $N \in \mathcal{Z}^+$. If (\ref{taylor}), (\ref{mean_cond}), and (\ref{remainder_cond}) are satisfied and $n=O(N)$, then 
    \begin{align}
 \M_n(\bm U_N) &= \E[\M(\bm U)] + O_p(n^{-1/2}),\\
  \sqrt{n}\left\{\M_{n}(\bm U_N) - \E[\M(\bm U)] \right \} &\overset{d}{\to} N(\bm 0, \Sig_M), \text{ if } \lim_{n\to\infty}(n/N) = 0,
  \end{align}
  where $\Sig_M$ is some positive definite matrix.
\end{theorem}

\noindent Theorem \ref{consistency1} implies that the sample estimator for $\mathcal{S}_{Y|\X}$ is $\sqrt{n}$ consistent for estimating the true central space from the aggregate of the marginal central spaces of $\mathcal{S}_{u^\top\D_n|\X}$ if $N\to\infty$  at a rate faster than $n$. Therefore, for any random response object $Y$ in a compact metric space $(\Omega_Y, d)$, Proposition \ref{surr_sdr} and Theorem \ref{consistency1} guarantees the recovery of a consistent unbiased estimator for $\spc$, i.e., $\mathrm{span}\big(\M_n(\bm U_N)\big) \subseteq \spc$.

\section{Simulation Experiments}
In this section, we evaluate the performance of our proposed surrogate-assisted methods compared to other competing methods in the literature where possible on synthetic data. For each method, we measure the estimation accuracy as the Frobenius norm of the difference in projection matrices of the true and estimated bases. This evaluation metric is common in the SDR literature. 

Let $\B$ be the true basis of $\spc$ and $\hat \B$ be the estimate of $\B$. Define the accuracy measure $\Delta$ as
\begin{align}
	\Delta = \lVert \bP_{\B} - \bP_{\hat \B} \rVert_F,
\end{align}  
where $\bP_\A = \A(\A^\top \A)^{-1}\A^\top$ and $\lVert.\rVert_F$ is the matrix Frobenius norm. Thus, smaller values of $\Delta$ indicate more accurate estimates.

\subsection{Euclidean response}
Fix $p \in \{10,20\}$ and $n \in \{100,500\}$ and generate the predictor as $\X_i \sim t_5(\bm 0, \I_p), i=1,\ldots,n$. %, where $\Sig_{ij}=\rho^{|i-j|}$ for $\rho \in [0,1]$. 
Let $\beta_1^\top = (1,1,0,\ldots, 0) \in \R^p$ and generate the response as follows:
\begin{enumerate}[label=\Roman*.]
    \item $Y_i = (1 + \beta_1^\top\X_i)\epsilon_i$
    \item $Y_i = \cfrac{0.5 X_{1i}^3}{0.5+(1.5+X_{2i})^2} + \epsilon_i$,
\end{enumerate}
where $\epsilon_i \overset{i.i.d.}{\sim} N(0,1)$.\\
Model I is a heteroscedastic error single-index model with $\B = \beta_1$. In Model II, $\B = (b_1, b_2)$ where $b_1=(1,0,\ldots,0)^\top$ and $b_2=(0,1,0,\ldots,0)^\top$. These models are popular in other SDR papers with the only exception being that we allow $\X$ to contain outliers instead of generating it from a multivariate normal distribution. We compare the classical OLS and SIR with their surrogate-assisted counterparts, sa-OLS and sa-SIR, respectively. For both SIR and sa-SIR, the number of slices is fixed at 5 under all settings. The simulation results are provided in Table \ref{scalar_sim}. 

In model I, because $\cms = \{\bm 0\}$, classical OLS is expected to perform poorly. Unsurprisingly, this handicap is not shared by sa-OLS as it performs much better than both the classical SIR and sa-SIR in small samples. In large samples however, sa-SIR outperforms the rest. All estimators improved significantly when $n$ is increased except the classical OLS. This is because regardless of the sample size, OLS will always yield $\bm 0$ as the estimate by design. For model II, OLS does not apply as it can only recover a single direction. However, sa-OLS is not confined to only single-index models and performs best in small samples. We see the same pattern in large samples as in Model I.

As expected, all estimators show consistency in large samples for fixed $p$ but show less accuracy with an increase in $p$ when we fix $n$, under all settings. Overall, the results in Table \ref{scalar_sim} demonstrate the superiority or at least noninferiority of the surrogate technique even in the Euclidean space. 

\begin{table}[htb]
\centering
\caption{Mean (standard deviation) of $\Delta$ based on 500 random samples.}
\label{scalar_sim}
\resizebox{\textwidth}{!}{
\begin{tabular}{ccccccc}
\hline
Model                & n                    & p  & OLS                 & sa-OLS          & SIR             & sa-SIR          \\ \hline
\multirow{4}{*}{I}   & \multirow{2}{*}{100} & 10 & 1.2804(0.0073) & 0.9996(0.0113)	& 1.0869(0.0107) & 1.0716(0.0114)\\
 &   & 20 & 1.3500(0.0037)	& 1.1573(0.008)	& 1.2392(0.0065)	& 1.2640(0.007)\\ \cline{2-7} 
 & \multirow{2}{*}{500} & 10 & 1.2635(0.0077) & 0.6893(0.0111) &	0.6172(0.0070) & 0.5691(0.0069) \\
&   & 20 & 1.3413(0.0042) & 0.8697(0.0095) & 0.829(0.0068)	& 0.7783(0.0070)   \\ \hline
\multirow{4}{*}{III} & \multirow{2}{*}{100} & 10 & \multirow{2}{*}{NA} & 1.3773(0.0064) & 1.4499(0.0067)	& 1.5338(0.0084)\\
 &   & 20 &   & 1.5712(0.0045) & 1.6484(0.0044)	& 1.7757(0.0052) \\ \cline{2-7} 
 & \multirow{2}{*}{500} & 10 & \multirow{2}{*}{NA} & 1.1693(0.0081) & 1.1555(0.0093) & 1.0143(0.0087)\\
 &   & 20 &   & 1.3279(0.0048) & 1.3339(0.006) & 1.2973(0.0068) \\ \hline
\end{tabular}
}
\end{table}

\quad

\subsection{Distribution as response}
We consider four univariate distributions including mixed, discrete, skewed, and symmetric distributions. First, fix $p \in \{10,20\}$ and $n \in \{100,500\}$, and generate $\X_i \sim \ t_5(\bm 0, \I_p), i=1,\ldots,n$. Let $\beta_1^\top = (1,1,0,\ldots, 0)$ and $\beta_2^\top = (0,0,1,1,0\ldots, 0)$. Then generate response distributions $Y_i$ as a sample in $\R^{100}$ as follows:

\begin{enumerate}[label=\Roman*.]
	\item $Y_i \in \R^{100} \overset{i.i.d.}{\sim} 0.6N(\beta_1^\top\X_i, 1) + 0.4N(1, 2)$.
	\item $Y_i \in \R^{100} \overset{i.i.d.}{\sim} Poisson\big(e^{\beta_2^\top\X_i}\big)$.
        \item $Y_i \in \R^{100} \overset{i.i.d.}{\sim} Gamma\big(\alpha_Y(\X_i), \beta_Y(\X_i)\big)$,  where $\alpha_Y(\X_i) = \lfloor e^{\beta_1^\top\X_i} \rfloor $ and \\ $\beta_Y(\X_i) = e^{0.5\beta_2^\top\X_i}$.
        \item $Y_i \in \R^{100} \overset{i.i.d.}{\sim} N\big(\beta_1^\top\X_i, (\beta_2^\top\X_i)^2\big)$.
\end{enumerate}
Here, we compare the kernel Fr\'ech\'et OLS and SIR of \cite{zhang2023dimension}, which we denote respectively as kf-OLS and kf-SIR, with their respective surrogate-assisted counterparts. As before, we fix the number of slices to 5 for both kf-SIR and sa-SIR under all settings. The results are provided in Table \ref{distri_sim}.

Across all settings, we see from Table \ref{distri_sim} that the surrogate-assisted estimators outperform their Fr\'ech\'et kernel counterparts. For the Gaussian distributions, sa-OLS dominates the rest while sa-SIR leads the way for the skewed distributions, i.e., Poisson and Gamma. This is not surprising as the Gaussian kernel on which the kf-OLS and kf-SIR are based are known to be susceptible and vulnerable in the presence of outliers likely present in models II and III.

\begin{table}[htb]
\centering
\caption{Mean (standard deviation) of $\Delta$ based on 500 random samples.}
\label{distri_sim}
\resizebox{\textwidth}{!}{
\begin{tabular}{ccccccc}
\hline
Model & $n$  & $p$  & kf-OLS & sa-OLS & kf-SIR  & sa-SIR          \\ \hline
\multirow{4}{*}{I}   & \multirow{2}{*}{100} & 10 & 0.2760(0.0123) & 0.2355(0.0077)	 & 0.4084(0.0066)	& 0.3513(0.0057) \\
&   & 20 & 0.4346(0.0150)	& 0.3673(0.0097)	& 0.6461(0.0097) & 0.5576(0.0086) \\ \cline{2-7} 
 & \multirow{2}{*}{500} & 10 & 0.1483(0.0094) &	0.1094(0.0021) & 0.2039(0.0031) &  0.1800(0.0029)\\
&  & 20 & 0.2211(0.0123) &	0.1584(0.0037) & 0.2816(0.003) &  0.2482(0.0029)\\ \hline

\multirow{4}{*}{II}  & \multirow{2}{*}{100} & 10 & 1.0052(0.0095) & 0.8670(0.0108) & 0.9191(0.0208)	&  0.3238(0.0066)\\
&   & 20 & 1.1884(0.0066) &	1.0737(0.0085) & 1.1102(0.0171)	&  0.4989(0.0077) \\ \cline{2-7} 
 & \multirow{2}{*}{500} & 10 &  1.0092(0.0093) & 0.8632(0.0112)  & 1.1827(0.0162)	&  0.1589(0.0028)\\
&   & 20 & 1.1696(0.0066) &	1.0383(0.0094) & 1.2911(0.0124)	&  0.2273(0.0050) \\ \hline

\multirow{4}{*}{III} & \multirow{2}{*}{100} & 10 & 1.4431(0.0059) & 1.4101(0.0060) & 1.5567(0.0104)	& 1.1212(0.0110)\\
&   & 20 &  1.5841(0.0047) & 1.5384(0.0046) & 1.7057(0.0087) & 1.3965(0.0056)\\ \cline{2-7} 
& \multirow{2}{*}{500} & 10 & 1.4473(0.0062) & 1.3994(0.0062)  & 1.7211(0.0078)	&  0.5869(0.0108) \\
&    & 20 &  1.5788(0.0053) & 1.5091(0.0056) & 1.8362(0.0064) &  0.8331(0.0104) \\ \hline

\multirow{4}{*}{IV}  & \multirow{2}{*}{100} & 10 & 1.5198(0.0107) & 1.2278(0.0111) & 1.4155(0.0092)	&  1.3377(0.0073)\\
 &   & 20 & 1.7451(0.0073) & 1.4315(0.0096) & 1.6152(0.0081)	&  1.4787(0.005)\\ \cline{2-7} 
 & \multirow{2}{*}{500} & 10 &  1.4991(0.0109) &	1.1890(0.0096) & 1.3471(0.0097) & 1.2893(0.0074) \\
&  & 20 &  1.7400(0.0070)	& 1.3075(0.0059) & 1.4535(0.0089) &	 1.3709(0.0043) \\ \hline
\end{tabular}
}
\end{table}

\subsection{Network as response}
As before, we fix $p \in \{10,20\}$ and $n \in \{100,500\}$ and generate the predictor as $\X_i \sim t_5(\bm 0, \I_p), i=1,\ldots,n$. We consider a classical random network model (Erd\"os-R\'enyi) and networks based on the stochastic block model (SBM). For all models, we fix the network size $N_V=20$ nodes. Each response network is generated from an adjacency matrix $\A \in \R^{20\times 20}$. 

In general, a stochastic block model network is generated as follows.
First, assume the $N_V$ nodes belong to $K$ classes and let $\M \in \R^{K\times K}$ be a symmetric block matrix whose entries are $m_{kq}$ for $k,q = 1, \ldots, K$. Also let $\g_j$ be the block membership indicator of node $j$ with $$\g_j \overset{i.i.d.}{\sim} Multinomial (1, \balpha), \text{ where } P(g_{jk}=1)=\alpha_k \text{ and } \displaystyle\sum_{k=1}^K\alpha_k = 1.$$ Next, for $v_j$ in class $k$ and node $v_\ell$ in class $q$, 
  \begin{align}
      \A_{j\ell}|g_{jk}, g_{\ell q} \overset{i.i.d.}{\sim} Bernoulli(P_{j\ell}),  
      \label{sbm}
  \end{align}
where $P_{j\ell}$ is is the probability of a link between $v_j$ in block $k$ and $v_\ell$ in block $q$, for $j,\ell \in 1,\ldots,N_V$. 

For this experiment, we consider four networks based on the models that follow.
\begin{enumerate}[label=\Roman*.]
	\item $Y_i \sim \A$ with
           $\A_{jk} = \begin{cases} 1 \overset{i.i.d.}{\sim} Bernoulli\big(plogis(-1.5+0.5\beta_1^\top\X_i)\big), \\
           0, \text{ otherwise},
           \end{cases}$ \\
      where $j,k = 1,\ldots, N_V$ and  $plogis(.)$ is the cumulative distribution function (CDF) of the logistic distribution.    
      
For models II-IV, we fix $K=3$ and $\balpha=(0.4,0.3,0.3)^\top$.

  \item $Y_i \sim \A$ with $P_{j\ell} = plogis\big(\mu_{j\ell} + \beta_1^\top\X_i\big)$, where
  \begin{align*}
      \mu_{j\ell}=\g_{j}^\top logit(\M)\g_{\ell} \text{ and } \M = \begin{pmatrix}
      0.45 & 0.05 & 0.05\\
      0.05 & 0.45 & 0.05 \\
      0.05 & 0.05 & 0.45
  \end{pmatrix}
  \end{align*}

  \item $Y_i \sim \A$ with $P_{j\ell} = ppois\big(\log(\bmu_{j\ell}) + \beta_1^\top\X_i\big)$, where 
  \begin{align*}
      \bmu_{j\ell}=\g_{j}^\top \M \g_{\ell}, \
      \M = \begin{pmatrix}
      6 & 1 & 1\\
      1 & 6 & 1 \\
      1 & 1 & 6
  \end{pmatrix}, \text{ and } ppois(.) \text{ is the CDF of Poisson distribution}.
   \end{align*}
       
    \item $Y_i \sim \A$ with $P_{j\ell} = \Phi\big(\bmu_{j\ell}+ (\beta_1^\top\X_i)^3\big)$, where $\Phi(.)$ is the CDF of the normal distribution. The mean and variance of the distribution are $\bmu_{j\ell}=\g_{j}^\top \M \g_{\ell}$ and $\sigma^2=1$, respectively. We set $\M = 2\I_3$.
\end{enumerate}

\begin{table}[htb!]
\centering
\caption{Mean (standard deviation) of $\Delta$ based on 500 random samples.}
\label{network_sim}
\resizebox{\textwidth}{!}{
\begin{tabular}{ccccccc}
\hline
Model                & n                    & p  & sa-OLS (CD)          & sa-OLS (DD)          & sa-SIR (CD)          & sa-SIR (DD)          \\ \hline
\multirow{4}{*}{I}   & \multirow{2}{*}{100} & 10 & 0.4900 (0.006)  & 0.2848 (0.0038) & 0.3021 (0.0046) & 0.2805 (0.0044) \\
                     &                      & 20 & 0.6808 (0.0059) & 0.4182 (0.0042) & 0.4793 (0.0066) & 0.4538 (0.0068) \\ \cline{2-7} 
                     & \multirow{2}{*}{500} & 10 & 0.2680 (0.0031) & 0.1467 (0.0021) & 0.1516 (0.0024) & 0.1435 (0.0023) \\
                     &                      & 20 & 0.3753 (0.0030) & 0.2050 (0.0022) & 0.2107 (0.0024) & 0.1991 (0.0024) \\ \hline
\multirow{4}{*}{II}  & \multirow{2}{*}{100} & 10 & 0.5920 (0.0066) & 0.5064 (0.0058) & 0.5104 (0.0060) & 0.4982 (0.0059) \\
                     &                      & 20 & 0.8337 (0.0057) & 0.7448 (0.0053) & 0.8132 (0.0079) & 0.7958 (0.0078) \\ \cline{2-7} 
                     & \multirow{2}{*}{500} & 10 & 0.2815 (0.0031) & 0.2384 (0.0027) & 0.2282 (0.0028) & 0.2245 (0.0028) \\
                     &                      & 20 & 0.4049 (0.0032) & 0.3451 (0.0027) & 0.3305 (0.0027) & 0.3256 (0.0027) \\ \hline
\multirow{4}{*}{III} & \multirow{2}{*}{100} & 10 & 0.5244 (0.0058) & 0.4446 (0.0050) & 0.4280 (0.0053) & 0.4194 (0.0052) \\
                     &                      & 20 & 0.7456 (0.0055) & 0.6494 (0.0049) & 0.6598 (0.0070) & 0.6475 (0.0069) \\ \cline{2-7} 
                     & \multirow{2}{*}{500} & 10 & 0.2556 (0.0029) & 0.2156 (0.0027) & 0.2018 (0.0026) & 0.2006 (0.0026) \\
                     &                      & 20 & 0.3615 (0.0030) & 0.3067 (0.0027) & 0.2873 (0.0027) & 0.2857 (0.0027) \\ \hline
\multirow{4}{*}{IV}  & \multirow{2}{*}{100} & 10 & 0.3364 (0.0046) & 0.2895 (0.0044) & 0.2826 (0.0045) & 0.2768 (0.0045) \\
                     &                      & 20 & 0.4853 (0.0046) & 0.4126 (0.0045) & 0.4259 (0.0054) & 0.4208 (0.0056) \\ \cline{2-7} 
                     & \multirow{2}{*}{500} & 10 & 0.1717 (0.0024) & 0.1526 (0.0024) & 0.1448 (0.0025) & 0.1425 (0.0025) \\
                     &                      & 20 & 0.2400 (0.0024) & 0.2115 (0.0024) & 0.2004 (0.0024) & 0.1973 (0.0024) \\ \hline
\end{tabular}
}
\end{table}

Although we used unweighted graphs in models I-IV, it appears the graph diffusion distance (DD), which is based on the edges is a better way to capture information about the predictors that influence the links than the degree centrality distance regardless of how the network was generated.

\subsection{Functional response}
Fix $p \in \{10,20\}$ and $n \in \{100,500\}$, and generate the fixed predictor $\X_i \sim t_5(\bm 0, \I_p), i=1,\ldots,n$. Next, generate 30 random time points $t\in\R^{30} \overset{i.i.d.}{\sim} Unif(0, 10)$. Then generate time-varying intercept $\alpha(t) = 2\sin(\pi + \pi t/5)$. Let $\beta_1^\top = (1,1,0,\ldots, 0)$ and $\beta_2^\top = (0,0,1,1,0,\ldots,0)$. The response trajectories $Y_i(t)$ are generated as follows:
\begin{enumerate}[label=\Roman*.]
	\item $Y_i(t) = \alpha_0(t) + 2\sin\left(\pi t/2 + \beta_1^\top\X_{i}\right) + \epsilon_{i}(t)$,
      \item $Y_i(t) = \alpha_0(t) + \sin\left(\pi t/2 + \beta_1^\top\X_{i}\right) + \sin\left(\pi t/2 - \beta_2^\top\X_{i}\right)  + \epsilon_{i}(t)$,
      \item $Y_i(t) = \alpha_0(t) + \cos\left(\pi t/2 + \beta_1^\top\X_{i}\right) + \cos\left(\pi t/2 - \beta_2^\top\X_{i}\right)  + \epsilon_{i}(t)$,
      \item $Y_i(t) = \alpha_0(t) + 2\cos\left(\pi t/2 + \beta_1^\top\X_{i}\right) + \sin\left(\pi t/2 + \beta_2^\top\X_{i}\right)\epsilon_{i}(t)$,
\end{enumerate}
where $\epsilon_{i}(t) \overset{i.i.d.}{\sim} N(0,1)$. 

\begin{table}[htb]
\centering
\caption{Mean (standard deviation) of $\Delta$ based on 500 random samples.}
\label{func_sim}
\resizebox{0.8\textwidth}{!}{
\tiny
\begin{tabular}{ccccc}
\hline
Model                & n                    & p  & sa-OLS          & sa-SIR          \\ \hline
\multirow{4}{*}{I}   & \multirow{2}{*}{100} & 10 & 0.5651 (0.0094) & 0.6126 (0.0109) \\
                     &                      & 20 & 0.7590 (0.0100) & 0.9415 (0.0130) \\ \cline{2-5} 
                     & \multirow{2}{*}{500} & 10 & 0.3006 (0.0038) & 0.2970 (0.0038) \\
                     &                      & 20 & 0.4151 (0.0039) & 0.4085 (0.0038) \\ \hline
\multirow{4}{*}{II}  & \multirow{2}{*}{100} & 10 & 1.6362 (0.0077) & 1.6967 (0.0066) \\
                     &                      & 20 & 1.8227 (0.0047) & 1.8533 (0.0041) \\ \cline{2-5} 
                     & \multirow{2}{*}{500} & 10 & 1.3376 (0.0057) & 1.3485 (0.0051) \\
                     &                      & 20 & 1.4162 (0.0032) & 1.4212 (0.0029) \\ \hline
\multirow{4}{*}{III} & \multirow{2}{*}{100} & 10 & 1.3898 (0.0065) & 1.4326 (0.0055) \\
                     &                      & 20 & 1.5221 (0.0044) & 1.5602 (0.0044) \\ \cline{2-5} 
                     & \multirow{2}{*}{500} & 10 & 1.3338 (0.0057) & 1.3460 (0.0052) \\
                     &                      & 20 & 1.4174 (0.0031) & 1.4213 (0.0029) \\ \hline
\multirow{4}{*}{IV}  & \multirow{2}{*}{100} & 10 & 1.4194 (0.0051) & 1.4343 (0.0051) \\
                     &                      & 20 & 1.5313 (0.0039) & 1.5403 (0.0044) \\ \cline{2-5} 
                     & \multirow{2}{*}{500} & 10 & 1.3573 (0.0046) & 1.3601 (0.0044) \\
                     &                      & 20 & 1.4287 (0.0026) & 1.4280 (0.0027) \\ \hline
\end{tabular}
}
\end{table}

\noindent As seen in Table \ref{func_sim}, sa-OLS and sa-SIR are very similar for the most part. However, sa-OLS appears to be slightly better than sa-SIR, especially in small samples.
%\newpage

\section{Real data analysis}
\subsection{COVID-19 transmissions in the U.S.}
The Coronavirus Disease 2019 (COVID-19) pandemic has severely impacted health systems and countries around the world. The U.S. in particular has seen very high case transmissions and mortality associated with the virus since December 2019. In this analysis, we focus our attention on the State of Texas, one of the largest and most populous states in the U.S. 

\subsubsection{Data}
Our data contains information about the daily COVID-19 transmission rate per 100,000 persons in each of the 254 counties across the State of Texas reported between 08/1/2021 and 02/21/2022. This data is available on the website of the Centers for Disease Control and Prevention (\cite{coviddata}). The second part of the data consists of county demographic information from the the 2020 American Community Survey (\cite{acs2020}). 

\begin{figure}[htb]
	\includegraphics[width=5.8in,height=3.1in]{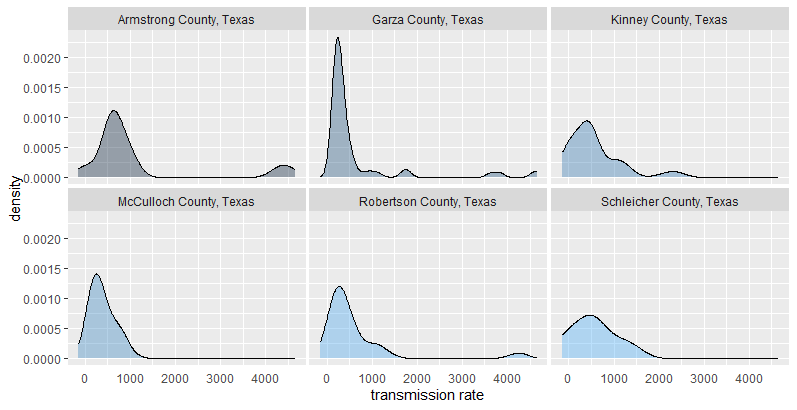}
	\caption{Distributions of COVID-19 case transmission rate across counties in the State of Texas between 08/1/2021 and 02/21/2022.}
	\label{distributions}
\end{figure}

The transmission rate is calculated as the change in the 7-day moving average of newly reported COVID-19 cases for a given county. Thus, for each county we have multiple reported rates ranging between 24 and 189 during the study period. In this analysis, we take the distribution of the transmission rates in the study period for each county as the response. The distributions of six select counties are given in Figure \ref{distributions}. We see from Figure \ref{distributions} that the distributions are all right-skewed with the highest peak at about 500 case transmissions per 100k persons. The shape of the distributions also show that it will be difficult to adequately describe the transmission distributions using one or two statistics such as the mean and variance.

We select eight county demographics including the percentage of people who are: Non-Hispanic Black, Hispanic, adults aged 65 and above, adults who have less than high school education, people who live below the poverty line, renter-occupied homes, and people on public assistance. We believe these predictors remained fairly constant over the study period and were less likely to be affected by local policy implementations such as mask mandates.

%Next, we smoothed the histograms of the transmission rates for each county to obtain probability density functions. This was done using the {\it Fr\'echet package} in $R$. 

\subsubsection{Analysis}
First, all predictors were marginally standardized. Then we estimated the bases of the central space for regressing the distributions of transmission rates on the demographic variables using the surrogate-assisted OLS and SIR and their kernel Fr\'ech\'et counterparts. We determined the structural dimension $d$ to be 1 based on the leading eigenvalues of the candidate matrices for all four methods. The estimated bases are given in Table \ref{dist_bases}. 

\begin{table}[htb!]
\centering
 \caption{estimated bases based of $\spc$}
\begin{tabular}{lcccc}
\hline
& \multicolumn{1}{l}{kf-OLS} & \multicolumn{1}{l}{sa-OLS} & \multicolumn{1}{l}{kf-SIR} & \multicolumn{1}{l}{sa-SIR} \\ \hline
\% Non-Hispanic Blacks  & 0.0612 & 0.1728 & 0.6439 & 0.4389 \\
\% Hispanics & -0.2771 & -0.3974 & 1.0861  & -0.3751 \\
\% Adults 65+  & -0.7254 & -0.8304 & 0.0988 & -0.6190 \\
\% No high school diploma & 0.3848 & 0.3949 & -0.3065 & -0.0339 \\
\% Living below poverty line & 0.2294  & 0.0284 & -0.7800 & -0.2963 \\
\% Unemployed & -0.0225 & 0.0455 & -0.1506 & 0.1828 \\
\% Renter-occupied homes & -0.8263 & -0.6710 &  0.2339  & 0.0091 \\
\% On public assistance & 0.0359 & 0.0969 & 0.5934 & 0.1939 \\
\hline
\end{tabular}
\label{dist_bases}
\end{table}

To compare the consistency of the estimates, we find the stability measure 
$$\Delta_s = \lVert \bP_{\hat\B_C} - \bP_{\hat \B_R}\rVert_F,$$ 
where $\hat\B_C$ is the estimate based on the complete sample and $\hat\B_R$ is the estimate based on a reduced random sample, which is 60\% of the full sample. 
\begin{table}[htb]
    \centering
    \caption{Mean (standard deviation) of $\Delta_s$ based on 500 random samples.}
    \begin{tabular}{ccccc}
    \hline
    & kf-OLS  & sa-OLS & kf-SIR  & sa-SIR \\ \hline
 & 0.7328 (0.0224) & 0.6487 (0.0185) & 1.1574 (0.0104) & 0.5409 (0.0080) \\
\hline
    \end{tabular}
    \label{consistency}
\end{table} 

\noindent From the results in Table \ref{consistency}, the surrogate-assisted estimates appear to be more stable compared to their Fr\'echet counterparts with sa-SIR being the most reliable. Based on the sa-SIR estimates in Table \ref{dist_bases}, it appears the percentage of adults aged 65 years and above has the most influence on the transmission distribution followed by the percentage of racial minorities living in the county. The percentage of people living in the county also have some moderate influence on the distribution of transmissions. These findings appear to be consistent with the literature on who is severely impacted by the COVID-19.

\begin{figure}[htb]
\includegraphics[width=\textwidth,]{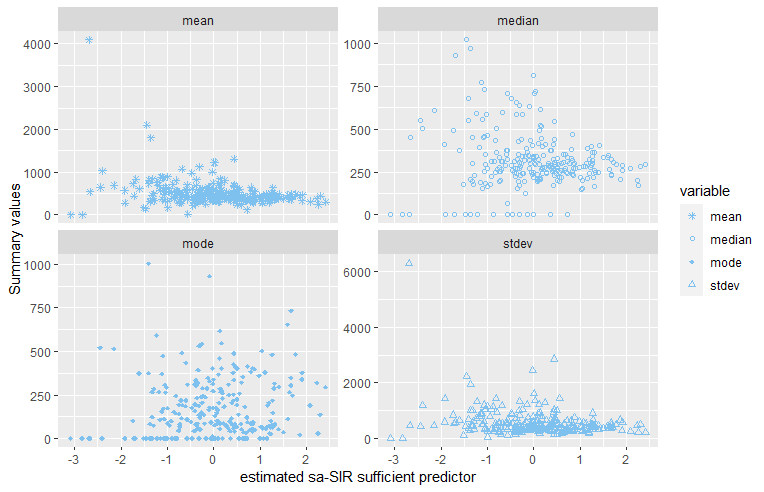}
	\caption{Scatter plot of some summary statistics of the distributions of COVID-19 transmission rates vs the estimated sufficient predictor based on sa-SIR estimator.}
	\label{centaltendency1}
\end{figure}

To gain more insights into the relationship between the transmission distributions and the demographics, we explore the relationship between different aspects of the distributions and the estimated sufficient predictor $(\hat\B^\top\X)$. The plots of four summary statistics: mean, median, mode and standard deviation of transmission rates for all counties versus the first sufficient predictor based on sa-SIR are given in Figure \ref{centaltendency1}. 

The mean and standard deviations do not appear to have a strong relationship with the predictors and also suggest the presence of outliers in the data. On the other hand, the median and mode appear to have a heteroscedastic linear relationship with the sufficient predictor. They also suggest the presence of outliers in the data. The implication of these plots is that an analysis based on mean transmission rates as response would likely find no relationship with the demographics although other measures of the center such as median and mode of the transmissions do appear to be associated with the demographics. This highlights one of the importance of using distributions as response rather than a point estimate of repeated measurements.

\subsection{Human brain structural connectivity}
The human brain is a complex organ that forms part of the central nervous system, which controls several functions in the body. Recent technologies such as magnetic resonance imaging (MRI) has allowed us to capture the structure and activities of the brain. This analysis focuses on the association between predictors including age, weight, and height and the structural connectivity network of the human brain.

\subsubsection{Data}
Our data is publicly available on the Open Science Framework (OSF) at \href{https://osf.io/yw5vf/}{https://osf.io/yw5vf/} and published in \cite{vskoch2022human}. The data contains the connectivity matrices, demographic, and clinical information of 88 healthy individuals. We consider the weighted adjacency matrix among 90 cortical regions of interests (ROI) as our response of interest.  Each matrix entry represents a porportion of tractography streamlines between any two ROIs. The predictors considered are age, weight, and heights of the individual subjects in the study. The details about how the data was generated and compiled can be found in \cite{vskoch2022human}. 

\begin{figure}[htb]
	\includegraphics[width=5.8in,height=3.1in]{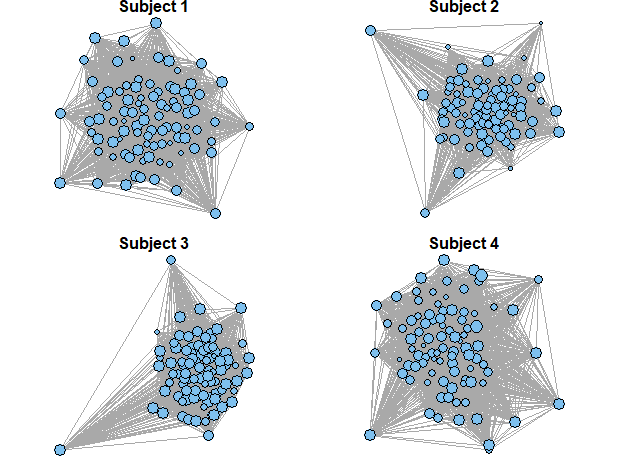}
	\caption{Brain connectivity networks of the first four subjects in the data set. Node size is proportional to the degree of the node.}
	\label{brain_networks}
\end{figure}

The networks of the first fours subjects in the data set are provided in Figure \ref{brain_networks}. Just by observation, the networks appear to be highly connect and differ by subject.\\

\subsubsection{Analysis}
We first standardize the predictors and then apply sa-OLS and sa-SIR estimators based on the degree centrality distances (CD) and diffusion distances (DD) of the weighted adjacency matrices. The results are provided in Table \ref{brain_bases}.\\

\begin{table}[htb!]
\centering
\caption{Estimated bases of $\spc$}
\begin{tabular}{lcccc}
\hline
       & sa-OLS (CD) & sa-OLS (DD) & sa-SIR (CD) & sa-SIR (DD) \\ \hline
Age    & 0.0744      & -0.2524     & 0.5273      & -0.2984     \\
Weight & -0.5527     & -0.5243     & -1.2538     & -0.5389     \\
Height & 1.3227      & -0.5447     & 1.0566      & -0.5196     \\ \hline
\end{tabular}
\label{brain_bases}
\end{table}

Again, for consistency, the stability measure 
$$\Delta_s = \lVert \bP_{\hat\B_C} - \bP_{\hat \B_R}\rVert_F,$$ 
where $\hat\B_C$ is the estimate based on the complete sample and $\hat\B_R$ is the estimate based on a reduced random sample (60\% of the full sample) is given in Table \ref{consistency2}. 

\begin{table}[htb!]
    \centering
    \caption{Mean (standard deviation) of $\Delta_s$ based on 500 random samples.}
    \begin{tabular}{ccccc}
    \hline
    & sa-OLS (CD) & sa-OLS (DD) & sa-SIR (CD)  & sa-SIR (DD) \\ \hline
    & 0.0784 (0.0017) & 0.7520 (0.0147) & 0.6995 (0.0191) & 1.1652 (0.0138) \\
     \hline
    \end{tabular}
    \label{consistency2}
\end{table} 

\begin{figure}[htb]
\includegraphics[width=\textwidth,]{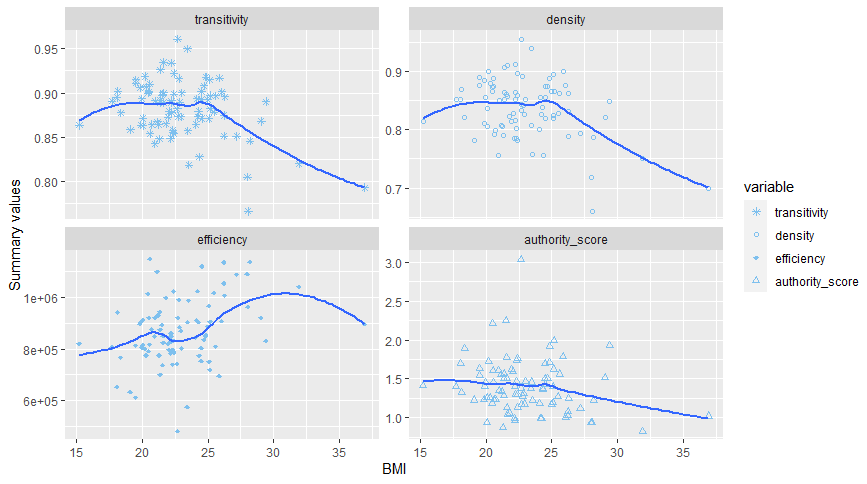}
	\caption{Scatter plot of some network properties vs BMI are provided. The blue line represents the loess curve.}
	\label{network_summaries}
\end{figure}

All methods consistently suggests that weight and height have the most influence on the structural connectivity of the brain compared to age. Therefore, we proceed to explore the relationship between some global network properties and body mass index (BMI), a measure which combines height and weight. We examine the relationship between four network properties: transitivity, density, efficiency, and authority score in Figure \ref{network_summaries}. 

Both transitivity and density are measures of connectivity. Transitivity reveals the existence of tightly connected communities, clusters, or subgroups while density measures the connectivity of the network. Figure \ref{network_summaries} shows that structural connectivity of the brain is fairly constant for BMI between 18 and 25 but declines for subjects with BMI exceeding 25. Efficiency gives a measure of reliability of communication and data transfer in the network. It appears from Figure \ref{network_summaries} that the brain network efficiency increases with BMI up to a BMI value of about 29.9, the borderline for overweight and obese, and then begin to decline gradually. Lastly, the authority score gives an indication of the number of hub points in the network and it appears to decrease with increase in BMI. These findings are consistent with other studies related to obesity and brain functionality.

%\newpage
\section{Discussion} 
This paper proposes a data visualization and dimension reduction technique for regressing metric space-valued responses on Euclidean predictors. Adopting the Euclidean embedding technique allows us to bypass the generation of high-dimensional universal kernels on the response space as proposed in \cite{zhang2023dimension}. Although we only illustrated this idea for responses that are Euclidean, univariate distributions, networks, and functional trajectories, our proposal readily extends to response objects in other metric spaces. Regardless of the response type, the key is to use the appropriate metric to get the pairwise distances. In addition, the proposed method can be extended to other SDR methods such as the minimum average variance estimate (MAVE) and outer product of gradient (OPG) and their extensions (\cite{xia2002mave, xia2009adaptive}). 

As demonstrated in both the simulation experiments and real data analyses, sa-OLS and sa-SIR are very powerful preprocessing tools in regression settings with complex responses. These real analyses also make a strong case for dimension reduction methods that go beyond the Euclidean space as they give room to explore the relationship between different aspects of the response objects and predictors. For instance, here we were able to see how different measures of the central tendency of the COVID transmission distributions vary with the sufficient predictor. On the contrary, if we were using the classical SDR methods, we would only be able to model a summary statistic of the transmission rates, say, the mean as a function of the demographics. But then, we may not gain any useful insights into how other statistics are influenced by the same factors without fitting models with those statistics. Hence, with the distributional responses we can kill several birds with one stone. 

Lastly, while being able to model a random response object as a function of Euclidean predictors expands our boundaries of applications, there is still room for more novel methods that can handle complex mix of responses and predictors in several metric spaces. Finally, the methods illustrated in this paper cover only linear sufficient dimension reduction. An extension to nonlinear SDR would be an interesting domain to explore in the future. 

\section*{Acknowledgements}
We thank the Children's Environmental Health Initiative at the University of Illinois Chicago for sharing their organized data on COVID-19 transmissions and the 2020 American Community Survey data.
% \bibliographystyle{elsarticle-harv}
% \bibliography{ref}

%\newpage
\appendix
\section{Appendix of proofs}

\begin{proof}[Proof of Theorem \ref{JL}]
An extensive proof of the Johnson–Lindenstrauss Lemma can be found in \cite{dasgupta2003elementary} and it is thus omitted.
\end{proof}

\begin{proof}[Proof of Proposition \ref{surr_sdr}] \quad \\
Let $\mathcal{S}$ denote a sufficient dimension reduction space of the regression of $Y$ versus $\X$ and $\mathfrak{F}$ be the collection of all subspaces for $Y$ versus $\X$. Given a measurable function $\psi$, let $\mathfrak{G}$ be the collection of all subspaces for $\psi(Y)$ versus $\X$. By definition, $Y\indep \X | \B^\top\X \Rightarrow \psi(Y) \indep \X | \B^\top\X$, which implies $\mathfrak{F} \subseteq \mathfrak{G}$. Therefore, $\bigcap\{\mathcal{S}: \mathcal{S}  \in \mathfrak{F} \} \supseteq \{\mathcal{S}: \mathcal{S}  \in \mathfrak{G} \}$.
%The proof follows directly from Theorem 2.3 of \cite{li2018sufficient} and is thus omitted.
\end{proof}

\begin{proof}[Proof of Proposition \ref{ensemble_sdr}] \quad \\
The proof of (a) and (b) follows directly from Lemma 2.1 while that of (c) follows from Theorem 2.1 in \cite{yin2011sufficient} and is thus omitted.
\end{proof}

\begin{proof}[Proof of Theorem \ref{consistency1}] \quad \\
    The details of the proof of this theorem can be seen directly from Theorems 3.2 and 3.3 in \cite{li2008projective}.
\end{proof}

 \bibliographystyle{agsm}
 \bibliography{ref}

\end{document}